# Breather decay into a vortex/anti-vortex pair in a Josephson Ladder

K. Segall[1], P. Williams[1], O. Svitelskiy[1], D. Edwards[1], N. Zhelev[1], G. Brummer[1] and J.J. Mazo[2]

[1]Physics and Astronomy Department, Colgate University, Hamilton, NY 13346, USA

[2]Departamento de Física de la Materia Condensada, Universidad de Zaragoza, 50009 Zaragoza, Spain

*Abstract: We present experimental evidence for a new behavior which involves discrete breathers and vortices in a Josephson Ladder. Breathers can be visualized as the creation and subsequent annihilation of vortex/anti-vortex pairs. An externally applied magnetic field breaks the vortex/anti-vortex symmetry and causes the breather to split apart. The motion of the vortex or anti-vortex creates multi-site breathers, which are always to one side or the other of the original breather depending on the sign of the applied field. This asymmetry in applied field is experimentally observed.* (PACS numbers: 05.45.Yv, 63.20.Pw, 74.50.+r, 74.81.Fa)

In complex nonlinear systems there can often be spatially or temporally coherent structures which emerge with marked particle-like properties [1] [2] [3] [4] [5]. Examples include solitons in nonlinear optics [6], kink dislocations in solids [7], skyrmions in magnetic materials [8] and vortices in superconductors [9] [10] or superfluids [11] [12]. Understanding these structures can be fundamental for many problems in physics and related fields. While many of these have been well-studied independently, how different types of structures within the same system interact and relate to each other is still very much an area of active research.

Arrays of superconducting Josephson junctions are excellent model systems to study such coherent structures [13] [14] [15]. They can be fabricated with adjustable parameters, easily scaled to large numbers and measured in a straightforward way. In addition, they are also inherently nonlinear due to the sinusoidal relationship between the phase of the superconducting wavefunction and the junction's supercurrent [16]. Two of the most fundamental coherent excitations in Josephson junction arrays are *Josephson vortices* and *discrete breathers*. Vortices are excitations which have spatially localized flux and an associated circulating current. They carry a topological charge. Discrete breathers [17], or more specifically rotational breathers or rotobreathers [18] [19] [20] [21], are time-periodic excitations which have spatially localized energy and no net topological charge. Of the different geometries of Josephson arrays, the *Josephson Ladder* has been demonstrated to support both vortices and rotobreathers in prior experiments.

In previous work [22] it has been noted that a rotobreather in a Josephson Ladder can be equivalently thought of as a time sequence of intermittent creation and subsequent annihilation of vortex/anti-vortex pairs. This stems from the fundamental relation between the vorticity, *n*, and the circulation of the superconducting phase gradient $\nabla\varphi$ around a plaquette in the ladder [23]:

$$\oint \nabla\varphi \cdot dl = 2\pi(n-f). \qquad (1)$$

Here *f* is the externally applied frustration. In a rotobreather, the phase difference across one or more of the junctions in a plaquette rotates in time. From equation (1), this necessitates a non-zero value of *n* at some point in the periodic breather solution. Typical breather solutions go from *n* = 1 (indicating a vortex) to *n* = -1 (anti-vortex) and back again in the plaquette neighboring the breather junction [22].

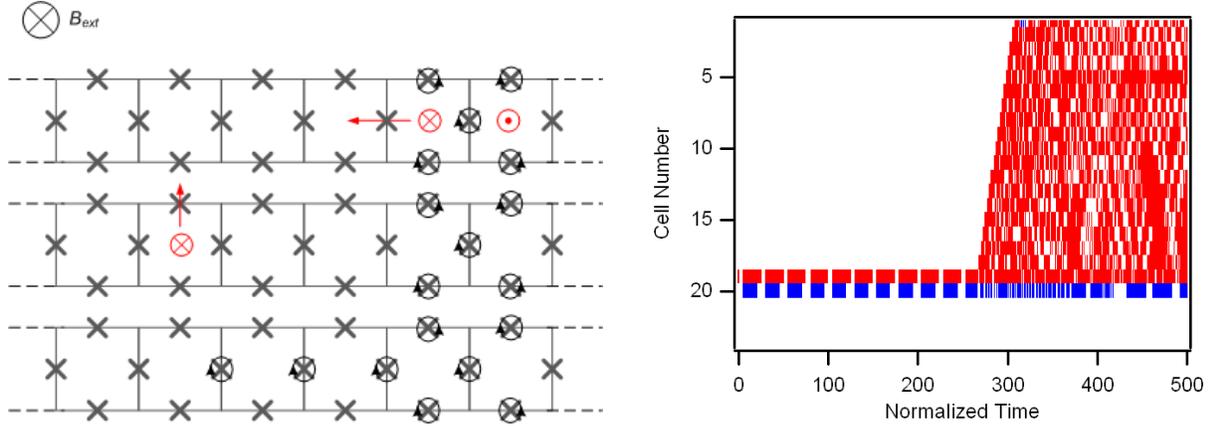

*Figure 1: (a) Dynamics of breather decay in a Josephson ladder. The "X"s indicate junctions, and those with a circle around them are in the voltage state with their phases rotating. The breather is shown at its point in the cycle where a vortex/anti-vortex pair has been created. The external magnetic field breaks the symmetry and allows the vortex to move to the left. After moving some number of cells it leaves the ladder. Junctions which have been passed by the vortex may be left in the voltage state, all of which are to the left of the breather. Under a negative applied field the anti-vortex would move to the right and the picture would be reversed. (b) Time-dependent simulation of the vorticity in our experiment. Cell number is on the left axis, time is on the bottom axis, and color indicates vorticity, with red indicating positive vorticity (vortex) and blue indicating negative vorticity (anti-vortex). The normalized value of current is I = 0.4 and the value of frustration is f = 0.06. At time 270 the vortex destabilizes and moves down the ladder, leaving that side of the ladder in the voltage state.*

Thus there exists a fundamental connection between breathers and vortices. Although this picture is correct mathematically, there has been no experimental observation to confirm it.

In this paper we show measurements of a new behavior in a Josephson Ladder where a rotobreather is destabilized and "split" into its composite vortex and anti-vortex under an applied magnetic field. The applied magnetic field breaks the vortex/anti-vortex symmetry allowing one or the other to separate from the breather and move down the array. The movement of the vortex or anti-vortex leaves some number of junctions in the voltage state, but only on one side of the breather. This result in an asymmetric distribution of switching currents as a function of applied magnetic field, which we have measured. Our work puts the before-mentioned connection between breathers and vortices on solid experimental footing. It also experimentally demonstrates a new decay mechanism for a discrete breather while adding to the ever-growing list of nonlinear effects in Josephson arrays.

The essential physics of our experiment is given in figure 1. A Josephson ladder is shown (fig. 1a) with an externally applied field that is into the page, defined to be positive. A vortex is indicated with its associated field into the page, while an anti-vortex is indicated with its field out of the page. The breather is shown at a point in its cycle where a vortex/anti-vortex pair has been created. The applied magnetic field breaks the vortex/anti-vortex symmetry. After the current reaches a certain value, the breather splits apart and the vortex moves to the left. When the vortex passes by a given junction, it causes a $2\pi$ rotation of the junction's phase, which may leave that junction in the voltage state. After moving a number of cells the vortex exits the ladder, leaving *m* junctions in the voltage state; in the case shown *m* = 3. Here the *m* junctions are consecutive, but that does not have to be the case. The dynamics of the vortex motion is too fast to be seen experimentally, but which junctions are left in the voltage state can be measured. In the case shown, all of the junctions in the voltage state are to the left of the breather. If the sign of the applied field is negative, then the picture will be flipped: the anti-vortex will move to the right and all junctions in the voltage state will be to the right of the breather.

Figure 1b shows a time-dependent simulation of this breather "splitting" in a Josephson ladder with parameters matched to the experiment (given below). To perform the simulations, we numerically integrated equations (5) – (8) in Trias et al. [22]  For the applied currents, we used the experimental protocol to generate a breather described later in the paper. The simulations were done without thermal noise. Shown in figure 1b is a color plot of the vorticity $n$, as defined in equation (1); red indicates a positive vorticity while blue indicates a negative vorticity. One can see that at the start of the simulation, the breather can be seen as a periodic creation and annihilation of a vortex/anti-vortex pair. At time = 270, the destabilization occurs and the vortex moves to the left to the end of the ladder. In the simulation, all of the junctions to the left of the breather are left in the voltage state. In the experiment, thermal noise and junction non-uniformity may result in only some of the junctions in the voltage state. However, all junctions in the voltage state will always be to one side or the other of the breather.

Our experiment consists of an anisotropic Josephson ladder with $N$ = 24 cells (24 vertical junctions, 23 top junctions and 23 bottom junctions). An electrical schematic is shown in figure 2a and an SEM micrograph is shown in figure 2b. The vertical junctions have an area of 10.75 µm$^2$ and a critical current of $I_{cv}$ = 6.3 µA; the full critical current of all 24 vertical junctions is 151 µA. The horizontal junctions have an area of 5.3 µm$^2$ and a critical current of $I_{ch}$ = 3.1 µA. The ladder anisotropy is $\eta = I_{cv}/I_{ch}$ = 0.49. The damping of the junctions ($\Gamma$), defined through $\Gamma^2 = \Phi_0/(2\pi I_c R_N^2 C)$, is equal to 0.044 for both horizontal and vertical junctions; here $I_c$ is the critical current, $R_N$ is the normal state resistance, and $C$ is the junction capacitance. The calculated geometrical inductance of each loop in the ladder is $L$ = 52 pH. This results in a coupling parameter of $\lambda = \Phi_0/(2\pi L I_{cv}) \approx 1$. The total current applied to the vertical junctions is denoted as the array current $I_a$; the current which is applied locally to one specific junction to create the breather is $I_{loc}$. On-chip series bias resistors (not shown) of about $R_b$ = 100 Ω attempt to distribute the array current equally.

Figure 2c shows the time-dependent currents that are applied to the array to create the breather. The array current is brought to a value of about 60 µA and held there, while the local current is spiked to about 10 µA to ensure that the breather junction is driven into the voltage state. The breather current is then reduced to zero. Following that the array current is ramped to about 200 µA to ensure that the whole ladder is switched into the voltage state. In our experiment we created a breather in junction 20, while measuring the voltage in junctions 19, 20, and 21. Figure 2d shows the time-dependence of the voltages of these junctions showing successful breather creation. Junction 20 is pushed into the voltage state as the local current is applied, and remains in a state with a voltage of about 1 mV after the local current is removed. Meanwhile the two neighboring junctions, 19 and 21, do not switch into the voltage state at that point, so a breather has been successfully created. The breather in junction 20 is in a resistive state and its voltage increases linearly with the current until it reaches about 1.3 mV, at which point it switches to the gap voltage of about 2.6 mV. At this point in time junctions 19 and 21 also jump to the voltage state. This procedure was repeated about 2000 times at each value of background field.

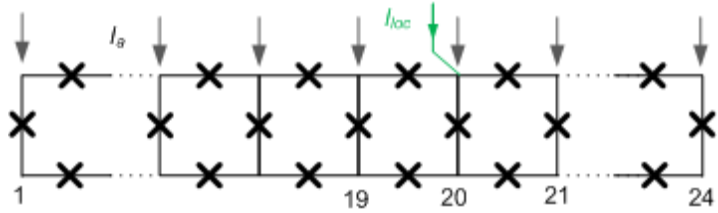
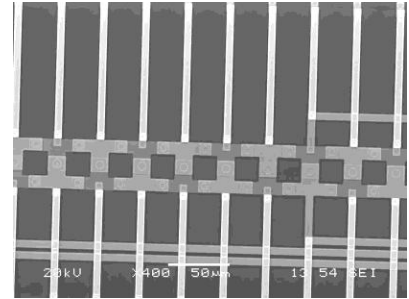
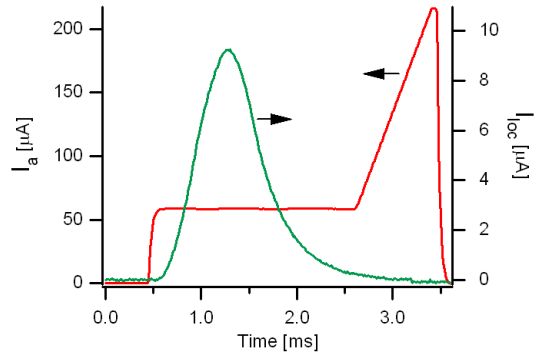
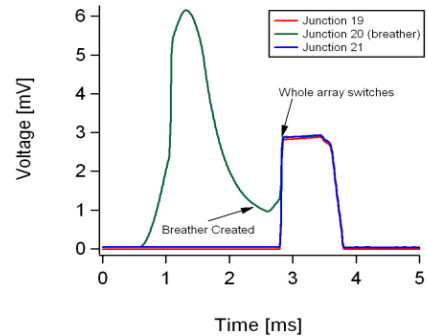

*Figure 2: Experimental breather creation. (a) Schematic of the ladder. The array current ($I_a$) is applied to all vertical junctions and the local current ($I_{loc}$) is applied only to junction 20. (b) Scanning electron micrograph of the Josephson ladder showing junctions 3-12. The lines running horizontal are voltage leads. (c) Time sequence of currents applied to the array to create the breather. The array current is brought to some value and held there while the local current is spiked and then returned to zero. (d) Voltage measurements in junctions 19, 20 and 21 showing creation of a breather near $f=0$.*

Breathers were successfully created about 2/3 of the time using this procedure; in the remaining cases the breather junction retrapped into the superconducting state once the local current was reduced. We refer to events where breathers were *not* created as type I events; events where breathers were created will be divided into type II events and type III events, as described below.

Figure 3 shows the two possible observed scenarios when we successfully create a breather. Instead of plotting voltage versus time, we now plot voltage versus current for junctions 19, 20 and 21. The curves are offset from each other for clarity. A type II event is shown in Fig. 3a. As the current is ramped, we can see junction 20 enters the voltage state at about 60 µA. At a current of about 100 µA all three junctions switch to the gap voltage and the ladder is in a homogenous whirling state. Type II events are distinguished by the fact that junctions 19 and 21 switch at the same current, the current when the whole ladder switches to the whirling state. Meanwhile, a type III event is shown in Fig. 3b. In the event shown, junction 19 switches into the voltage state at the same time as junction 20, so a multi-site breather of (at least) $m=2$ has been created at that value of field. At a higher current, junction 21 switches to the gap voltage along with the rest of the ladder. Events where junctions 19 and 21 switch at *different* currents are categorized as type III events. We define the switching current as the current when the voltage passes a threshold of 0.6 mV, as shown in the figure. In the case shown in Fig. 3b, junction 21 has a larger switching current than junction 19.

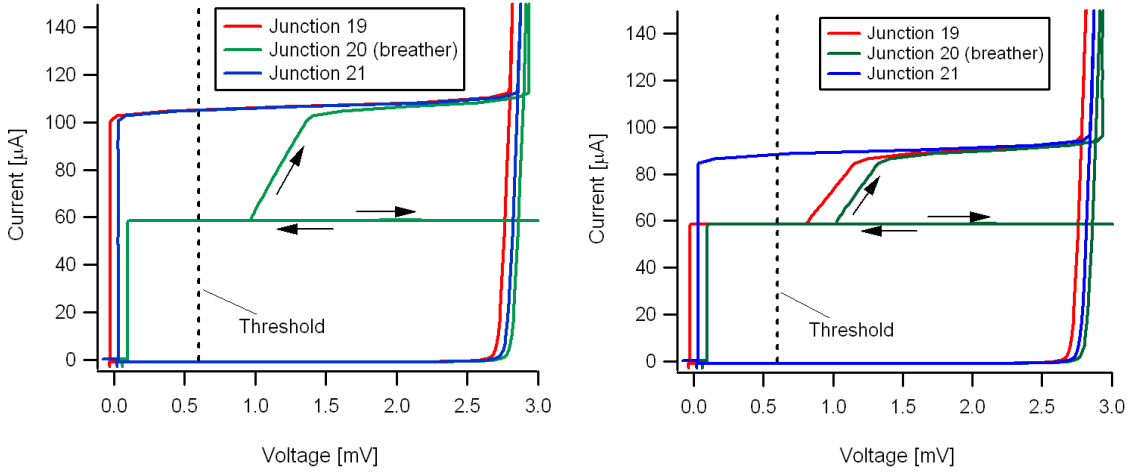

*Figure 3: Current-voltage curves for junctions 19, 20 and 21 showing a type II event (left) and a type III event (right) at a field of f = -0.175. Curves are offset for clarity. In the type II event, junctions 19 and 21 switch at the same time when the whole ladder goes to the whirling state. In the type III event, Junction 19 switches into a resistive state at roughly the same time as the breather junction, while junction 21 is in the superconducting state until the whole ladder switches.*

Besides the differences in switching currents between junctions 19 and 21, information can also be obtained by the slope of the IV curve in junction 20 after it goes into the breather state. Figure 4a shows 100 different I-V curves of junction 20 at $f$ = -0.19. One can see events with a range of slopes in the resistive state; we identified 8 different slopes in the data shown. These different slopes represent events with different number of junctions in the voltage state, i.e. events with different values of $m$. The slope varies because as the number of junctions in the voltage state changes, the current distribution of the array is slightly altered, since junctions in the resistive state will receive less current that those in the superconducting state. Although the series bias resistors try to even out these differences, the bias resistors cannot be made large enough to ensure a perfectly even current distribution. This point has been explored in previous work. The slope of the I-V curve ($dI_a/dV_b$) for a junction in an $m$-site breather is known [22] to be given as:

$$dI_a/dV_b = N/R_V [1 + 2\eta/ms + R_V/sR_b](1 - m/N), \quad (2)$$

where $R_V$ is the resistance of the junction in the breather state and $s$ is an integer indicating the type of breather [22]. Figure 4b shows the measured slopes versus $m$ and the fit using equation (2). For the fitting parameters, we used $N$ = 24, $R_b$ = 103 Ω, $R_V$ > 3000 Ω and $s$ = 2, indicating a top-bottom symmetric breather solution (if we try to use $s$ = 1 in Figure 5b we cannot fit the data with any reasonable choice of $R_b$ and $R_V$). $R_V$ is taken to be much larger than the normal state resistance (238 Ω) because breather voltages are in the subgap region; the fitting was not sensitive to the value of $R_V$ above a few kΩ. The only fitting parameter was $R_b$, which was designed to be 80 Ω but came out somewhat larger because of the Hypres etching process.[1] The value of $R_b$ was chosen to fit the largest slope, where $m=1$. We then assigned values for $m > 1$ to their closest predicted values of slope.

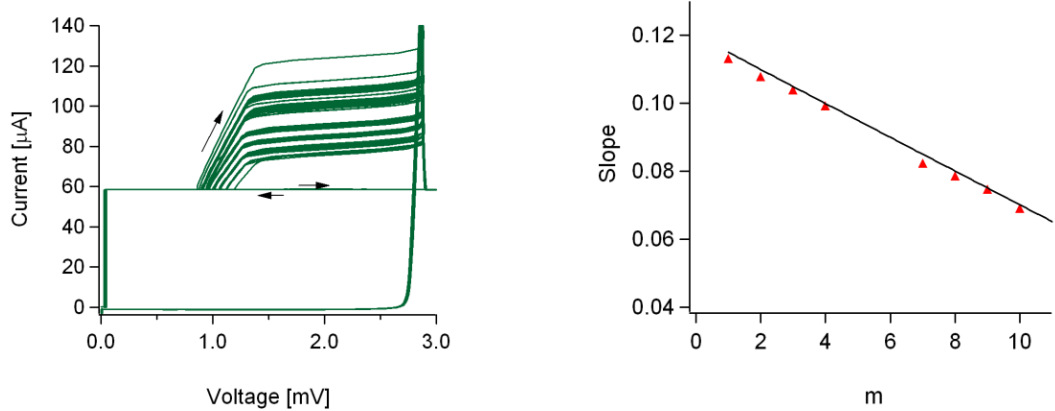

*Figure 4: (a) I-V characteristic of junction 20 for 100 sweeps. Different slopes are observed when the junction is in the voltage state. (b) Slope of the I-V curve versus the number of junctions in the multi-site breather. The line is a fit from equation (2).*

In Figure 4 we can also see clearly that the breather junction goes unstable at almost the same voltage each time: around 1.4 mV or about half of the gap voltage (2.8 mV). Previous experiments on the subgap currents of similar junctions [24] showed a large increase in the subgap current at this voltage, so this is not surprising. At these points, the whole ladder switches into the whirling mode. Since this happens at the same voltage, events with a larger slope or smaller $m$ will make it to higher switching currents before switching to the whirling mode. As we will see, this will separate the switching currents into discrete bands, with each band having a different value of $m$.

We now look at the full distribution of switching currents as a function of applied magnetic field, shown in Figure 5. First, in figure 5a, we plot the switching current versus magnetic field for the case of no breather in the ladder. Junction 19 is shown, but all junctions have the same behavior in this case. We see a periodic, SQUID-like modulation of the switching current with a triangular shape, similar to what was observed by the Ustinov group [25]. Near $f = 0.5$, we see a region of small switching current, due to the spontaneous creation of breathers; this also seen by the Ustinov group [26] in Josephson Ladders with very similar parameters.

In figures 5b and 5c we show the full switching current distribution of junctions 19 and 21, with a breather created in junction 20. A whole set of different new events are now seen. We label them into the three groups (I, II and III) mentioned previously. Events from group I switch at the largest currents and appear in both junctions 19 and 21. As previously mentioned, they represent failed breather creation, and follow the same magnetic field dependence as shown in figure 4a.

Events from group III are at the lowest switching currents and appear in one junction or the other, but not both. They correspond to the creation of a multi-site breather which includes either junction 19 or junction 21. Type III events from junction 19 indicate that junction 19 switches later than junction 21, so a multi-site breather was created which includes junction 21, which switched at about the same time as junction 20 (60 µA). Meanwhile, type III events from junction 21 correspond to events where junction 21 switches later than 19, so a multi-site breather was created which includes junction 19. (This was the case depicted in the I-V curves shown in figure 3b.) In figure 4d we histogram the *difference* between the switching current of junction 19 and junction 21 (junction 19 – junction 21). Here events that have a positive difference indicate that junction 19 switched later than 21, while events that have a negative difference indicate that junction 21 switched later than 19. As is clearly seen, positive events only occur when (*f* – Floor(*f*)) < 0.5, while negative events only occur when (*f* – Floor(*f*)) > 0.5.

---
[1] M. Radparvar, private communication

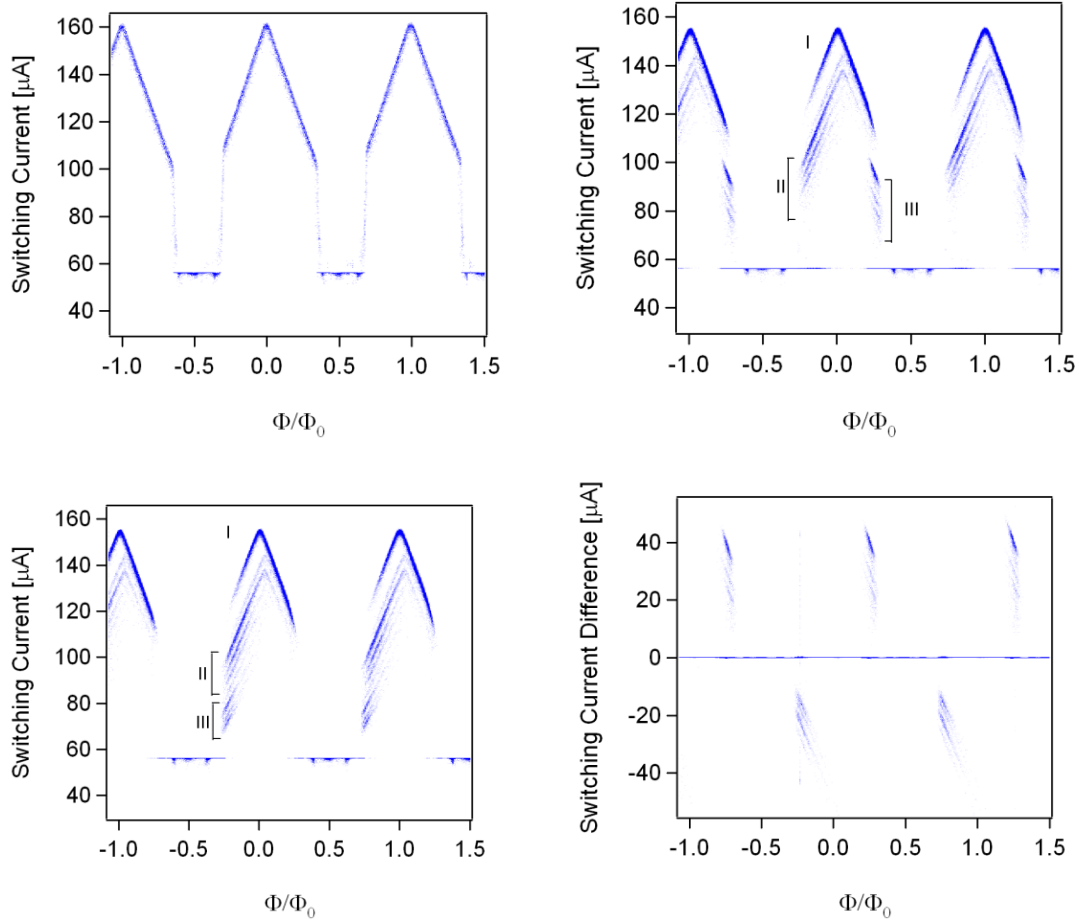

*Figure 5: Switch current versus frustration (f = $\Phi/\Phi_0$) for various scenarios. Dots indicate individual switching events. (a) Junction 19 with no breather in the ladder. (b) Junction 19 with a breather in junction 20. The three different sets of switching events are indicated. (c) Junction 21 with a breather in junction 20. (d) Switching current difference between junction 19 and 21. Positive currents indicate junction 19 switches after junction 21, while negative currents indicate junction 21 switches after junction 19.*

The type II events occur closer to *f* = 0 than the type III events, have larger switching currents, and mostly do not cause a difference in switching currents between junctions 19 and 21. The different values of *m* are what give the different "bands" of events in the switching current distributions. As mentioned before, the larger switching currents correspond to smaller values of *m*, while smaller switching currents correspond to larger values of *m*. Note that there are many more bands on the side where *f* < 0, which are events where the vortex spreads to the left, than on the side where *f* > 0, where the anti-vortex spreads to the right. Since there are more junctions to the left of junction 20 than to the right, more values of *m* are possible on the left. In fact we only see four bands when *f* > 0, which is what one would expect with four junctions to the right of junction 20. To the left of junction 20 we see *m* as large as 9 or 10. The asymmetry due to the vortex or anti-vortex motion takes on a different signature in the type II events, but still clearly indicates motion to one side or the other. For the type II events, junctions 19 and 21 mostly do not switch into the voltage state after the vortex or anti-vortex moves away from the breather; they switch together when the whole ladder reaches the whirling state. This possibility was seen in simulations with noise in earlier work [27].

In short, we have observed a new behavior in a Josephson ladder where one type of coherent excitation, a discrete breather, decays into two others, a vortex and anti-vortex, under an applied magnetic field. This "splitting" results in multi-site breathers created on one side of the ladder or the other, which can be ascertained through IV curve and switching current measurements. The number of junctions in the multi-site breather can be determined by measuring the slope of the I-V curve with involving the breather, and causes asymmetric bands of switching events. Future work will focus on better identifying the different possible states of multi-site breathers, both with experiments and numerical simulations.

We thank Daniel Schussheim for numerical simulation work and Dan Schult, Joe Amato and Terry Orlando for useful discussions. This work was supported by NSF DMR 1105444. JJM acknowledges financial support from Spanish MINECO through Project No. FIS2011-25167, co-financed by FEDER funds.